\documentclass[aps,prx,twocolumn,amsmath,amssymb,superscriptaddress,letterpaper]{revtex4}
\usepackage{times}
\usepackage{amsfonts}
\usepackage{mathrsfs}
\usepackage[version=3]{mhchem}
\usepackage{graphicx}
\usepackage{dcolumn}
\usepackage{bm}
\usepackage{color}

\usepackage[colorlinks,bookmarks=false,citecolor=blue,linkcolor=red,urlcolor=blue]{hyperref}
\def\be{\begin{equation}}       \def\ee{\end{equation}}
\def\bea{\begin{eqnarray}}      \def\eea{\end{eqnarray}}

\begin{document}

\title{Opposite-Mirror-Parity Scattering as the Origin of Superconductivity in Strained  Bilayer Nickelates}

\author{Congcong Le}
\thanks{These authors equally contributed to the work.}
\affiliation{Hefei National Laboratory, Hefei 230088, China}
\affiliation{RIKEN Center for Interdisciplinary Theoretical and Mathematical Sciences(iTHEMS), Wako, Saitama 351-0198, Japan}

\author{Jun Zhan}
\thanks{These authors equally contributed to the work.}
\affiliation{Beijing National Laboratory for Condensed Matter Physics and Institute of Physics, Chinese Academy of Sciences, Beijing 100190, China}
\affiliation{School of Physical Sciences, University of Chinese Academy of Sciences, Beijing 100190, China}

 \author{Xianxin Wu}\email{xxwu@itp.ac.cn}
 \affiliation{CAS Key Laboratory of Theoretical Physics, Institute of Theoretical Physics,
Chinese Academy of Sciences, Beijing 100190, China}

\author{Jiangping Hu}\email{jphu@iphy.ac.cn}
\affiliation{Beijing National Laboratory for Condensed Matter Physics and Institute of Physics, Chinese Academy of Sciences, Beijing 100190, China}
\affiliation{Kavli Institute for Theoretical Sciences, University of Chinese Academy of Sciences, Beijing 100190, China}
\affiliation{ New Cornerstone Science Laboratory, Beijing 100190, China}

\begin{abstract}
We study the electronic structure and doping-dependent instabilities of strained La$_3$Ni$_2$O$_7$ thin films using first-principles and functional renormalization group methods. We demonstrate that ordering tendencies are governed by Fermi surface scattering between electrons of opposite mirror parity. Under moderate hole doping, when the $d_{z^2}$ bonding band becomes incipient or crosses the Fermi level, robust $s_{\pm}$-wave superconductivity emerges from cooperative interlayer pairing reinforced by two competing spin-density-wave fluctuations. Compressive strain favors superconductivity in NiO$_2$ bilayers slightly away from the interface, whereas tensile strain induces pair-breaking nesting that suppresses pairing. Our results establish a unified microscopic scenario for superconductivity in pressurized bulk and strained thin-film nickelates, providing new insights into high-T$_c$ pairing in correlated quantum materials.
\end{abstract}

\maketitle

Despite decades of extensive research efforts into nickelates, the nickel age of superconductors only began in 2019 with the discovery of superconductivity in "infinite-layer" nickelates, specifically (Sr,Nd)NiO$_2$ thin films grown on a substrate \cite{Li2019,Osada2020,Pan2022,wang2022,ding_critical_2023}. Remarkably, four years later, nickelate superconductors entered the high-temperature regime akin to cuprates with the observation of superconductivity in a novel type of bilayer nickelate La\(_3\)Ni\(_2\)O\(_7\) (LNO) within the Ruddlesden-Popper phase under pressure, boasting an extraordinarily high transition temperature (T\(_c\)) of approximately 80 K \cite{sun2023}. Superconductivity emerges abruptly following a structural transition from the \textit{Amam} phase to the \textit{Fmmm}/\textit{I4mmm} phase as pressure increases. In contrast to the \(d^9\) electronic configuration of cuprates and "infinite-layer" nickelates, the Ni\(^{2.5+}\) in LNO exhibits a \(d^{7.5}\) configuration, with both \(d_{x^2-y^2}\) and \(d_{z^2}\) orbitals contributing to the low-energy electronic structure due to the presence of apical oxygens between layers~\cite{YaoDX,YZhang2023,Lechermann2023,Hirofumi2023possible,XWu,XJZhou2023,HHWen2023}. Additionally, the trilayer nickelate La$_4$Ni$_3$O$_{10}$ exhibits superconductivity under pressure with a lower T$_c$ of 20-30 K~\cite{Kuroki2023T,HHWen2024T,JZhao2023T,YQi2023T,MWang2023T}. Both bilayer and trilayer nickelates exhibit  spin density wave (SDW) orders~\cite{sun2023,HQYuan2023,HHWen2023,chen2024electronic,ren_resolving_2025}, showing close relation with superconductivity~\cite{XHChen2025-SDW}.
The pairing mechanism and symmetry are under intensive theoretical study~\cite{Wang327prb, lu2024interlayer,HYZhangtype2,XWu,FangYang327prl,WeiLi327prl,Hirofumi2023possible,YifengYang327prb,YifengYang327prb2,YiZhuangYouSMG,tian2023correlation,Dagotto327prb,zhang2024structural,Jiang_2024,PhysRevB.108.L201121,liao2023electron,ryee2024quenched,luo2023hightc,fan2023superconductivity,KuWeiprl,KJiang:17402,zhan2024cooperation,ChenHH2025,PhysRevB.111.144514,GuanGJ2025}. Theoretical calculations suggest that the pressure-driven Lifshitz transition, where the \(d_{z^2}\) bonding band crosses the Fermi level ($E_\text{F}$) to form a hole pocket, is crucial for bulk superconductivity~\cite{sun2023}. Spin fluctuations then promote an \(s_{\pm}\)-wave interlayer pairing~\cite{Hirofumi2023possible,XWu,Wang327prb,FangYang327prl}. Other studies emphasize interlayer exchange, Hund's coupling, and orbital hybridization in generating interlayer pairing in the strong-correlation limit~\cite{lu2024interlayer,HYZhangtype2,WeiLi327prl}. However, further experimental investigation into the electronic structures and pairing gaps in the bulk nickelates is challenged due to the high-pressure conditions. 

Intriguingly, recent studies have shown that compressively strained LNO thin films grown on the SrLaAlO$_4$ (SLAO) substrate exhibit superconductivity with a T\(_c\) exceeding 40 K at ambient pressure~\cite{ko2024signatures,zhou2024ambientpressuresuperconductivityonset40,liu2025superconductivitynormalstatetransportcompressively,bhatt2025resolvingstructuraloriginssuperconductivity}. Although this T\(_c\) is only about half the bulk value, these films provides a tunable and powerful platform for investigating the mechanisms underlying high-T\(_c\) superconductivity. First-principles calculations indicate that compressive strain eliminates the hole pocket around the M point, whereas tensile strain results in Fermi surfaces similar to those of the bulk material under pressure~\cite{Geisler2024,2024arXiv241204391Z}. This makes the absence of superconductivity in tensile-strained films~\cite{ko2024signatures,bhatt2025resolvingstructuraloriginssuperconductivity} a puzzling contradiction to the proposed bulk mechanism.  
Moreover, ARPES results on compressively strained LNO thin films are not yet consistent: one group reported heavy hole doping and a diffusive hole pocket around the M point~\cite{li2025photoemissionevidencemultiorbitalholedoping,yue2025correlated,2025arXiv250110409S}, another identified that the $d_{z^2}$ bonding band resides slightly below $E_\text{F}$~\cite{ShenZX2025}.
These findings motivate a comprehensive study of key ingredients for superconductivity in bilayer nickelates and prompt three questions: (1) What mechanism drives superconductivity in strained films, and how does it related to the $d_{z^2}$ bonding band? (2) Why do tensile-strained LNO thin films, despite hosting Fermi surfaces similar to the bulk, fail to exhibit superconductivity? (3) Do bulk and thin film LNO share the same pairing mechanism?

\begin{figure}[t]
\centerline{\includegraphics[width=0.45\textwidth]{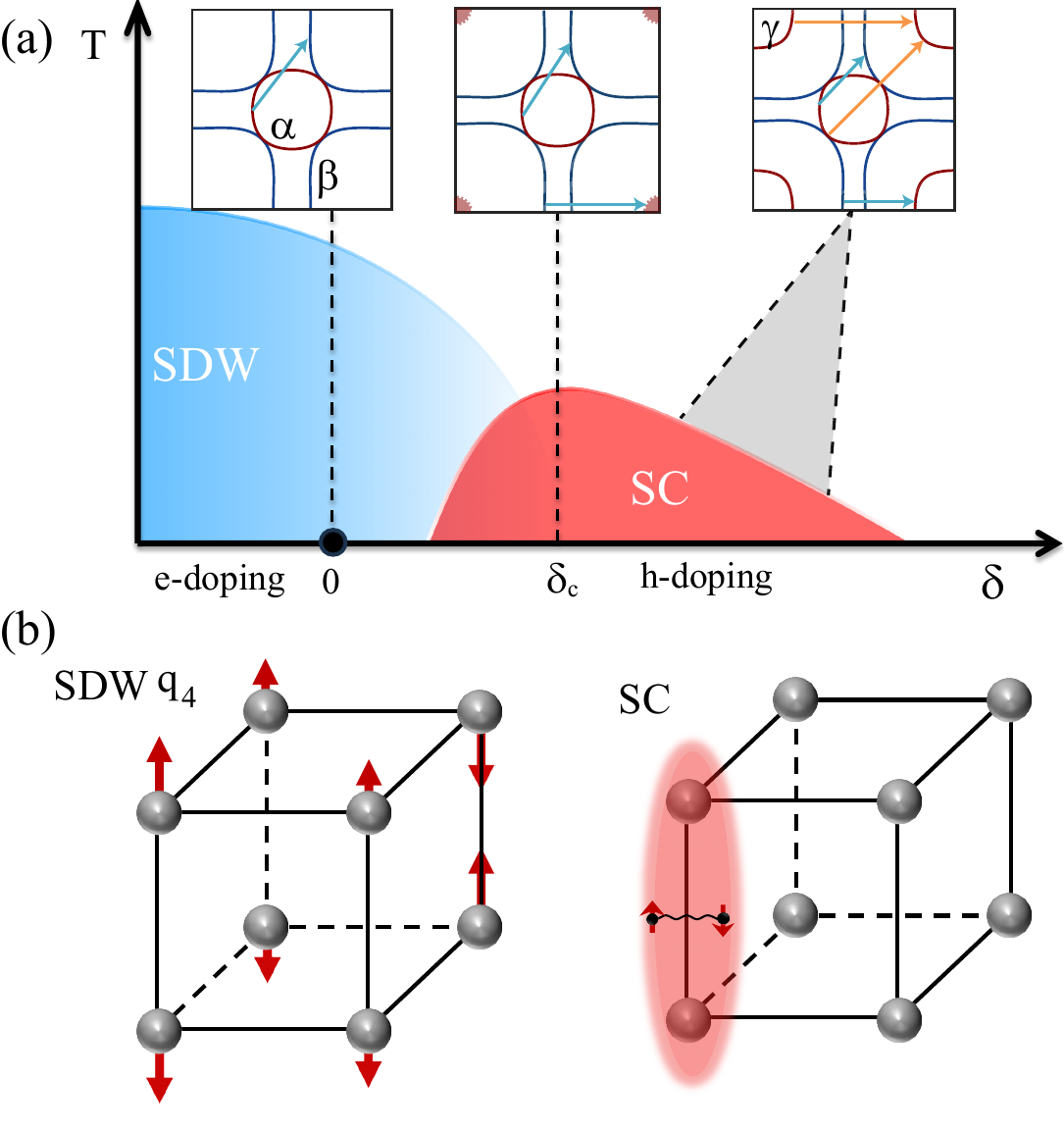}}
\caption{
 (a) Schematic temperature-doping phase diagram for compressively strained LNO thin films. Insets show three representative Fermi surfaces together with nesting vectors: blue arrows indicate nesting between bonding (red) and antibonding (blue) pockets, while the orange arrow indicates nesting between two bonding pockets. The incipient $d_{z^2}$ band is represented by the shaded sector around the corner and $\delta_c$ marks the Lifshitz transition. (b)  Schematics for the SDW with the wavevector of $\bm{q}_4$ and superconductivity with interlayer pairing.
 \label{fig0}}
\end{figure}

To address these questions, we examine the electronic structure and doping-dependent correlated instabilities of strained LNO thin films by employing first-principles and functional renormalization group (FRG) calculations. We uncover an intimate relation between ordering tendencies and Fermiology, summarized in Fig.~\ref{fig0}(a), where scattering between Fermi surfaces of opposite mirror parity (interlayer bonding/antibonding) plays an essential role in promoting correlated states. In compressively strained undoped films, nesting between the bonding \(\alpha\) and antibonding \(\beta\) pockets results an SDW with interlayer antiferromagnetic (AFM) configuration. With hole doping, the \(d_{z^2}\) bonding band moves toward $E_\text{F}$; once it becomes incipient, additional finite-energy spin fluctuations arise from nesting between the antibonding \(\beta\) pocket and the flat \(d_{z^2}\) bonding band. These two SDW fluctuations, with different in-plane wave vectors, closely compete and suppress the SDW instability. However, they cooperatively promote interlayer pairing, yielding robust \(s_{\pm}\)-wave superconductivity, as shown in Fig.~\ref{fig0}(b). When a bonding \(\gamma\) pocket fully appears, \(\beta\)–\(\gamma\) nesting enhances interlayer pairing, whereas \(\alpha\)–\(\gamma\) and \(\gamma\)–\(\gamma\) nestings generate ferromagnetic interlayer coupling and suppress interlayer pairing. The latter effect is stronger, causing $T_c$ to decrease and resulting a T$_c$ maximum near the Lifshitz transition. Our calculations suggest that superconductivity most likely originates from the NiO\(_2\) bilayer slightly away from the interface—where the \(d_{z^2}\) bonding band resides near $E_\text{F}$ and substrate-induced hole doping can occur—rather than from the interfacial bilayer. Under tensile strain, the \(\alpha\)–\(\gamma\) nesting is significantly enhanced, weakening interlayer pairing and driving an SDW instability despite the presence of the \(\gamma\) pocket. Finally, we discuss a unified scenario for superconductivity in both bulk and thin films and explore potential experimental implications.

\begin{figure}
\centerline{\includegraphics[width=0.45\textwidth]{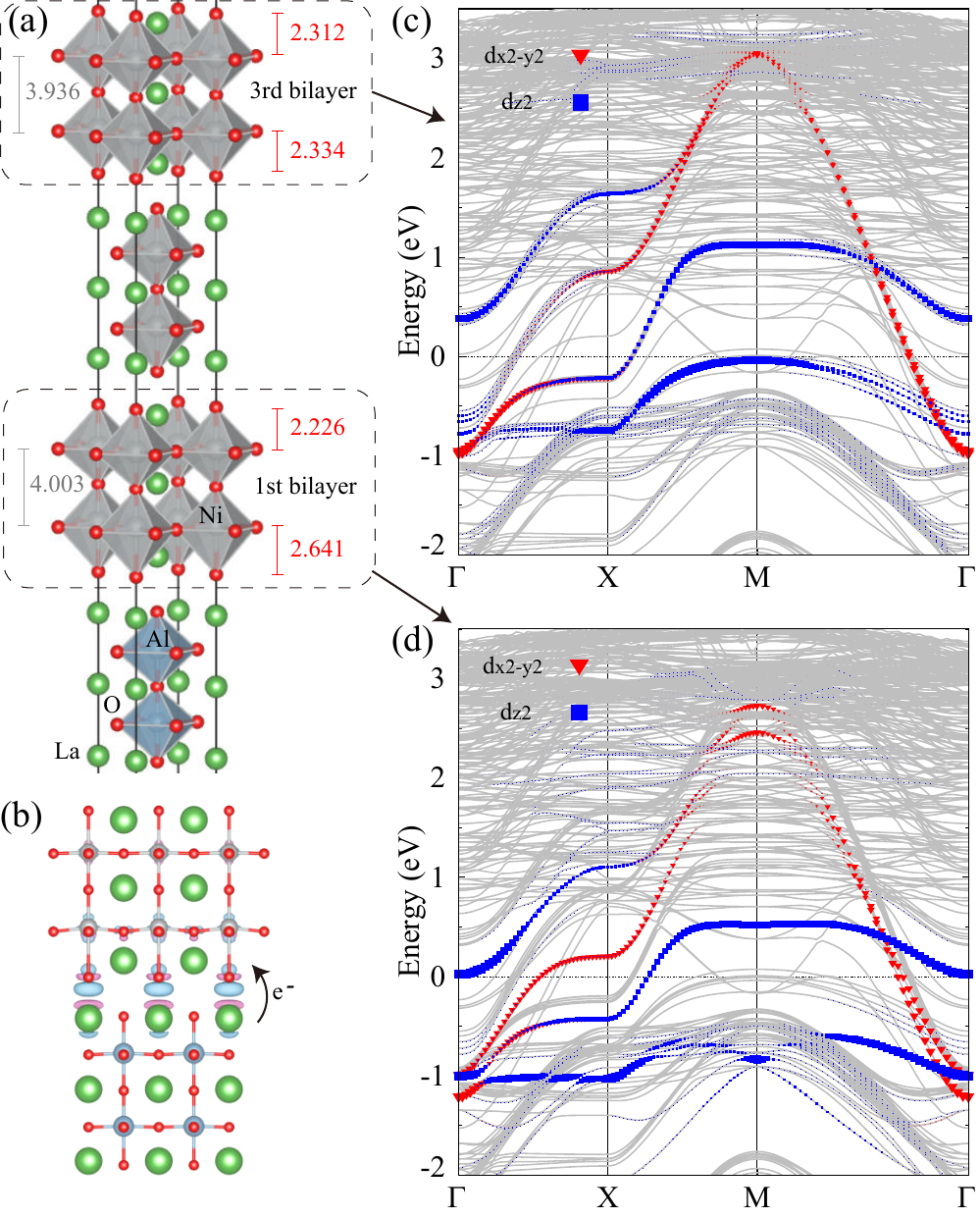}}
\caption{ (a) Part of optimized geometry of 3UC LNO on  \ce{SrLaAlO3} substrate, comprising six NiO$_2$ bilayers in total. (b) Charge density differences near the interface, where blue and red indicate charge accumulation and depletion, respectively. (c) and (d) {\it ab initio} orbital-resolved band structures for the 1st  and 3rd bilayers. \label{fig1}}
\end{figure}

{\it Electronic structures of LNO/SLAO interface}. We start with the crystal and electronic structures of three-unit-cell (3UC) pristine LNO on the SLAO substrate. The tetragonal LNO is assumed and initial interface is constructed according to the structure from scanning transmission electron microscope measurements~\cite{zhou2024ambientpressuresuperconductivityonset40,yue2025correlated} and further relaxation is done for the 3UC LNO and the first two layers of substrate. The details of first-principles calculations are shown in the supplementary materials (SM).
Fig.~\ref{fig1}(a) shows the optimized crystal structure, where only three bilayers of the film are shown along with part of the substrate, and  the complete structure is available in SM. The in-plane compressive strain generates an elongation along the c-axis for LNO, which mainly occurs on the LaO layers according to our calculations. To clearly demonstrate different structural changes at and near the interface, we select the first and third bilayers as representative examples and show their structural parameters. The out-of-plane distance between two Ni sites $d_{\text{Ni}-\text{Ni}}$ and outer apical Ni-O bond length $d_{\text{Ni}-\text{O}}$ within these bilayers are indicated in gray and red numbers, respectively. Across these bilayers, $d_{\text{Ni}-\text{Ni}}$ is approximately 4 \AA, close to the bulk value, and this value is slightly longer in the 1st bilayer than the 3rd bilayer due to the effect of substrate. Notably, the outer apical Ni-O distance is increased to 2.3 \AA~ away from the interface due to the elongation of LaO layers and $d_{\text{Ni}-\text{O}}$ in the 1st bilayer is elongated to 2.6 \AA. This elongation of the outer apical Ni-O bond results in a significant reduction in the onsite energy of the $d_{z^2}$ orbital in the 1st bilayer. Besides structural modifications, there is electrostatic doping from the substrate.  Fig. \ref{fig1}(b) displays charge density differences and blue and red indicate regions of charge accumulation and depletion, respectively, suggesting a net electron transfer from the substrate to the 1st bilayer.

Furthermore, the orbital-resolved band structures along high-symmetry paths for the 1st and 3rd bilayers are presented in Fig. \ref{fig1}(c) and (d), with a focus on the $e_g$ orbitals ($d_{x^2-y^2}$ and $d_{z^2}$). In the interfacial (1st) bilayer, the reduced onsite energy of the $d_{z^2}$ orbital induces a significant downward shift of its associated bands, as seen in Fig. \ref{fig1}(d). This shift causes the $d_{z^2}$ bonding
band to move away from $E_{\text{F}}$, while the $d_{z^2}$ anti-bonding states approach and make contact with $E_\text{F}$ at the $\Gamma$ point. The electron doping is approximately 0.34 without considering substrate-induced Sr diffusion. Consequently, the 1st bilayer hosts a single $\beta$ hole pocket around the M point from a mixture of $d_{x^2-y^2}$ and $d_{z^2}$ orbitals, and an $\alpha$ electron pocket around the $\Gamma$ point from the $d_{x^2-y^2}$ orbital. The asymmetry between two Ni sites splits the $d_{x^2-y^2}$ bands along the $\Gamma-\text{M}$ path. Incorporating interaction in the LDA+U formalism will induce an additional electron pocket around $\Gamma$ from the $d_{z^2}$ antibonding state (see SM). In contrast, the bilayer away from the interface, such as the 2nd and 3rd bilayers, show no charge doping and have band structure similar to bulk LNO at ambient pressure, as shown in Fig. \ref{fig1}(c). The $d_{z^2}$ bonding band just drops below $E_\text{F}$ ~\cite{2024arXiv241204391Z}, yielding an electron pocket around the $\Gamma$ point and a hole pocket around the M point. The substrate-induced Sr diffusion~\cite{zhou2024ambientpressuresuperconductivityonset40} will induce hole doping and make $d_{z^2}$ bonding band cross $E_\text{F}$ in these bilayers (see SM).

{\it Scenarios of superconductivity in compressed LNO thin films}.
Although superconductivity has been observed in strained LNO ultrathin films, its origin and which NiO\(_2\) bilayer hosts it remain elusive. Pinpointing the active bilayer is crucial for revealing the mechanism, as interfacial and interior bilayers have markedly different electronic structures. To investigate this issue, we construct effective tight-binding models by fitting hopping parameters to \textit{ab initio} band structures and study electronic instability by varying carrier doping and interaction parameters, based on FRG calculations. As discussed earlier, the Ni \(d_{x^2-y^2}\) and \(d_{z^2}\) orbitals dominate the low-energy physics in strained LNO thin films, similar to bulk LNO at both ambient and high pressures \cite{YaoDX,YZhang2023,Lechermann2023,Hirofumi2023possible,XWu,XJZhou2023,HHWen2023}. Therefore, a two-orbital tight-binding (TB) model can accurately describe the low-energy band structure \cite{YaoDX,YZhang2023,Lechermann2023,Hirofumi2023possible,XWu}, and the corresponding Hamiltonian is given by,
\begin{equation}
    \mathcal{H}_0=\sum_{ij,\alpha\beta,\sigma}t_{\alpha\beta}^{ij}c_{i\alpha\sigma}^\dagger c_{j\beta\sigma}-\mu\sum_{i\alpha\sigma} c_{i\alpha\sigma}^\dagger c_{j\alpha\sigma}.
\end{equation}
Here $i,j=(m,l)$ label the in-plane lattice site ($m$) and top/bottom layer index ($l=t,b$), $\sigma$ labels spin, and $\alpha,\beta=x,z$ represent the Ni orbitals with $x$ denoting the $d_{x^2-y^2}$ and $z$ the $d_{z^2}$ orbital. The chemical potential $\mu$ varies with charge doping. Details of the Hamiltonian matrix are provided in the SM. Compared to the bulk case, the asymmetry between the two Ni sites in the 1st bilayer at the interface breaks the mirror symmetry connecting them, resulting in slightly different onsite energies and in-plane hopping for the two layers. The bilayer slightly away from the interface preserves this mirror symmetry and shares the same TB model with the bulk, though with modified effective parameters. For the electronic interactions, we adopt the general multi-orbital Hubbard interactions,
\begin{equation}
    \begin{aligned}
    \mathcal{H}_{\mathrm{I}}&= \sum_{i\alpha}Un_{i\alpha\uparrow}n_{i\alpha\downarrow}+\sum_{i,\alpha\neq \beta}J_Pc_{i\alpha\uparrow}^{\dagger}c_{i\alpha\downarrow}^{\dagger}c_{i\beta\downarrow}c_{i\beta\uparrow}\\
    &+\sum_{i,\alpha< \beta,\sigma\sigma^{\prime}}(U^{\prime}n_{i\alpha\sigma}n_{i\beta\sigma^{\prime}}+J_Hc_{i\alpha\sigma}^{\dagger}c_{i\beta\sigma}c_{i\beta\sigma^{\prime}}^{\dagger}c_{i\alpha\sigma^{\prime}}),
    \end{aligned}
\end{equation}
where $U/U^{\prime}$ is the intra-/inter-orbital Hubbard repulsion, $J_{H}$ is the Hund ’s coupling, and
$J_P$ is the pair-hopping interaction. The standard Kanamori relations $U = U^{\prime} + 2J_{H}$ and $J_H=J_P$ are adopted in this work~\cite{Kanamori}. To resolve electronic instabilities, we employ the FRG approach, which treats all particle-hole and particle-particle channels on equal footing, providing an unbiased depiction of correlated states from weak to intermediate coupling regimes \cite{FWang2010,Metzner2012,Platt2013}. As the RG cutoff $\Lambda$ decreases, a divergent flow indicates an instability towards a symmetry-broken state in either particle-particle or particle-hole channel. Further details of the formalism are provided in the SM.

We begin by examining the electronic instability of the interfacial NiO\(_2\) bilayer. Compared to the pressurized bulk case, there is an increase in the weight of both the \(d_{x^2-y^2}\) bonding orbital on the \(\alpha\) pocket and the \(d_{z^2}\) antibonding orbital on the \(\beta\) pocket. Systematic calculations for both hole-doped ($\delta > 0$) and electron-doped ($\delta < 0$) cases across various Hund’s couplings show that the phase diagram is predominantly governed by a spin-density-wave (SDW) state associated with Fermi surface nesting between the \(\alpha\) and \(\beta\) pockets. In contrast, superconductivity only appears within a narrow range of weak Hund’s coupling with low critical cutoffs. These findings suggest that superconductivity is unlikely to originate from the interfacial NiO\(_2\) bilayer, consistent with robust superconductivity observed in samples with different interfaces~\cite{ko2024signatures,zhou2024ambientpressuresuperconductivityonset40,liu2025superconductivitynormalstatetransportcompressively}.

\begin{figure}[t]
\centerline{\includegraphics[width=0.48\textwidth]{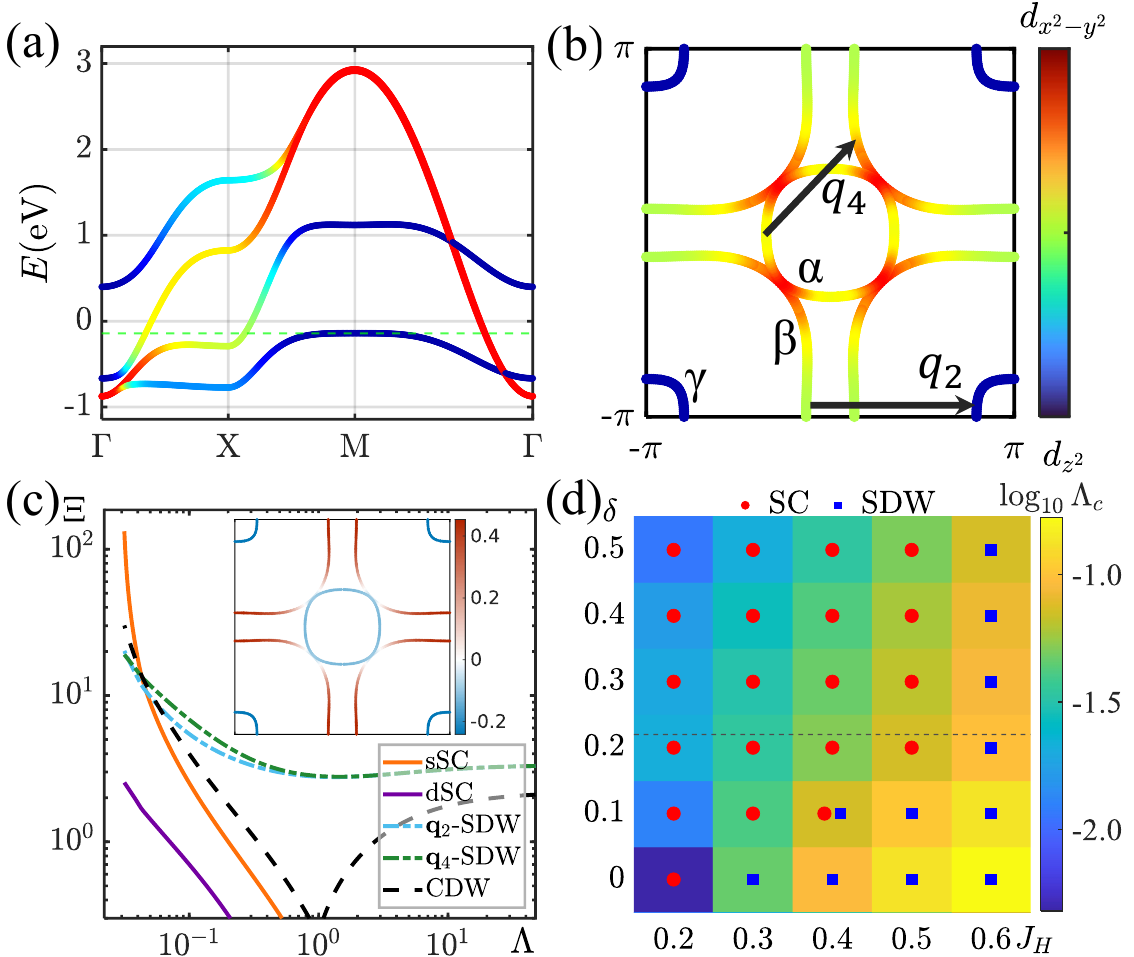}}
\caption{
(a) TB-model band structure of the third NiO$_2$ bilayer and corresponding Fermi surface (b) at the filling $n=2.7$ marked by the green dashed line in (a).  (c) Typical FRG flows of leading eigenvalues in SC, SDW and CDW channels with $U=3$ eV, $J=0.1U, \delta=0.3$. Superconductivity is leading with two SDW competing and sub leading and the corresponding superconducting gap is shown in the inset. (d) Phase diagram of instabilities as a function of Hund's coupling $J_H$ and hole doping $\delta$ at fixed $U=3$ eV. 
The color bar indicates the critical scale $\Lambda_c$ and the black dashed line denotes the Lifshitz transition point $\delta_c=0.22$.  \label{fig3}}
\end{figure}

Next, we investigate the instabilities in the NiO\(_2\) bilayer slightly away from the interface, which is influenced solely by epitaxial strain and Sr-diffusion-induced doping~\cite{zhou2024ambientpressuresuperconductivityonset40}. The orbital-resolved band structure is depicted in Fig.\ref{fig3}(a), showing two pockets in the absence of hole doping. The $d_{z^2}$ bonding band is located below $E_\text{F}$ due to its reduced onsite energy. However, with hole doping, this band approaches $E_\text{F}$, and eventually crosses $E_\text{F}$, resulting a Lifshitz transition when $\delta$ exceeds $\delta_c=0.22$.
Fig. \ref{fig3}(b) illustrates the Fermi surfaces in such a case with $\delta=0.3$. Our FRG calculations show that, in the undoped and slight-hole-doped cases, the system exhibits prominent SDW instability for strong Hund's coupling, while $s_{\pm}$ superconductivity occurs at weak Hund's coupling.
However, when the $d_{z^2}$-bonding band approaches $E_\text{F}$ (incipient) or crosses $E_\text{F}$ to generate a Lifshitz transition (black dashed line), superconductivity instability becomes enhanced and robust, and can even persist under strong Hund's coupling, as shown in Fig.~\ref{fig3}(d). After the Lifshitz transition, T$_c$ drops monotonically with increasing doping. A representative FRG flow for the case with a hole doping $\delta=0.3$ is displayed in Fig.~\ref{fig3}(c), where \(s_{\pm}\)-superconductivity is dominant, with the gap function depicted in the inset. This pairing state is primarily attributed to the interlayer pairing on $d_{z^2}$ orbital. 
These results suggest that when the $d_{z^2}$ bonding band becomes incipient or crosses $E_F$ to form the $\gamma$ pocket, it significantly suppresses SDW order and generates robust \(s_{\pm}\)-wave superconductivity, which is related to competing Fermi surface nesting.

This is intimately related to the interlayer magnetic coupling, which can be characterized by the bare interlayer susceptibility, defined as,
\begin{eqnarray}
&&\chi^{0}_{t\alpha, b\beta}(\bm{q},i\omega_n)=-\frac{1}{N}\sum_{\bm{k}\nu\nu'} a^{b\beta}_\nu(\bm{k})a^{t \alpha*}_{\nu}(\bm{k}) a^{t \alpha}_{\nu'}(\bm{k}+\bm{q}) \nonumber\\
&&\times a^{b\beta*}_{\nu'}(\bm{k}+\bm{q}) \frac{n_F[\epsilon_{\nu}(\bm{k})]-n_F[\epsilon_{\nu'}(\bm{k}+\bm{q})]}{i\omega_n+\epsilon_{\nu}(\bm{k})-\epsilon_{\nu'}(\bm{k}+\bm{q})}.
\label{chi}
\end{eqnarray}
Here $\nu/\nu'$ is the band index, $n_F(\epsilon)$ is the Fermi distribution function, $a^{l\alpha}_\nu(\bm{k})$ is the $l\alpha$-th component of the eigenvector for band $\nu$ by diagonalizing the TB model and $\epsilon_{\nu}(\bm{k})$ is the corresponding eigenvalue. Nesting between two bands with the same mirror parity, namely both bonding or antibonding, contributes the even-channel susceptibility $\chi^{0,e}$~\cite{Eremin_prb2024}. This yields a positive product of eigenvector elements in the above formula, resulting a positive $\chi^{0}_{t\alpha, b\beta}$ and, consequently, ferromagnetic interlayer coupling. In contrast, nesting between a bonding and an antibonding band contributes the odd-channel susceptibility $\chi^{0,o}$, and yields a negative product and a negative $\chi^{0}_{t\alpha, b\beta}$, leading to AFM interlayer coupling. When the $d_{z^2}$ band resides away from $E_F$, the nesting between bonding $\alpha$ and antibonding $\beta$ pockets contributes to significant SDW fluctuation at $\bm{q}_4$, inducing a dominant SDW instability. However, when the $d_{z^2}$-bonding flat band becomes incipient~\cite{ChenX2015,mishra2016s,Nakata2017} or crosses $E_F$ to generate the $\gamma$ pocket around the M point, this introduces an additional strong SDW fluctuation with the nesting vector $\bm{q}_2$ connecting $\beta$ and $\gamma$ pockets, competing with SDW fluctuation at $\bm{q}_4$. This is apparent from the FRG flow in the SDW channel shown in Fig.~\ref{fig3}(c), where the close competition between two SDWs with different wave vectors suppresses the SDW instability.
Remarkably, both Fermi surface nestings occur between bonding and anti-bonding bands and thus generate strong interlayer AFM fluctuations, collectively promoting interlayer pairing and stabilizing \(s_{\pm}\)-wave superconductivity. The presence of $\gamma$ pocket introduces strong interlayer FM coupling from $\gamma-\gamma$ nesting ~\cite{ryee2024quenched},  suppressing interlayer pairing and resulting a T$_c$ reduction with increasing doping.  
 Consequently, the observed superconductivity in compressed thin films is likely to occur in NiO\(_2\) bilayers slightly away from the interface, where the substrate can contribute to both strain effect and Sr-diffusion-induced hole doping~\cite{zhou2024ambientpressuresuperconductivityonset40}.

\begin{figure}[t]
\centerline{\includegraphics[width=0.48\textwidth]{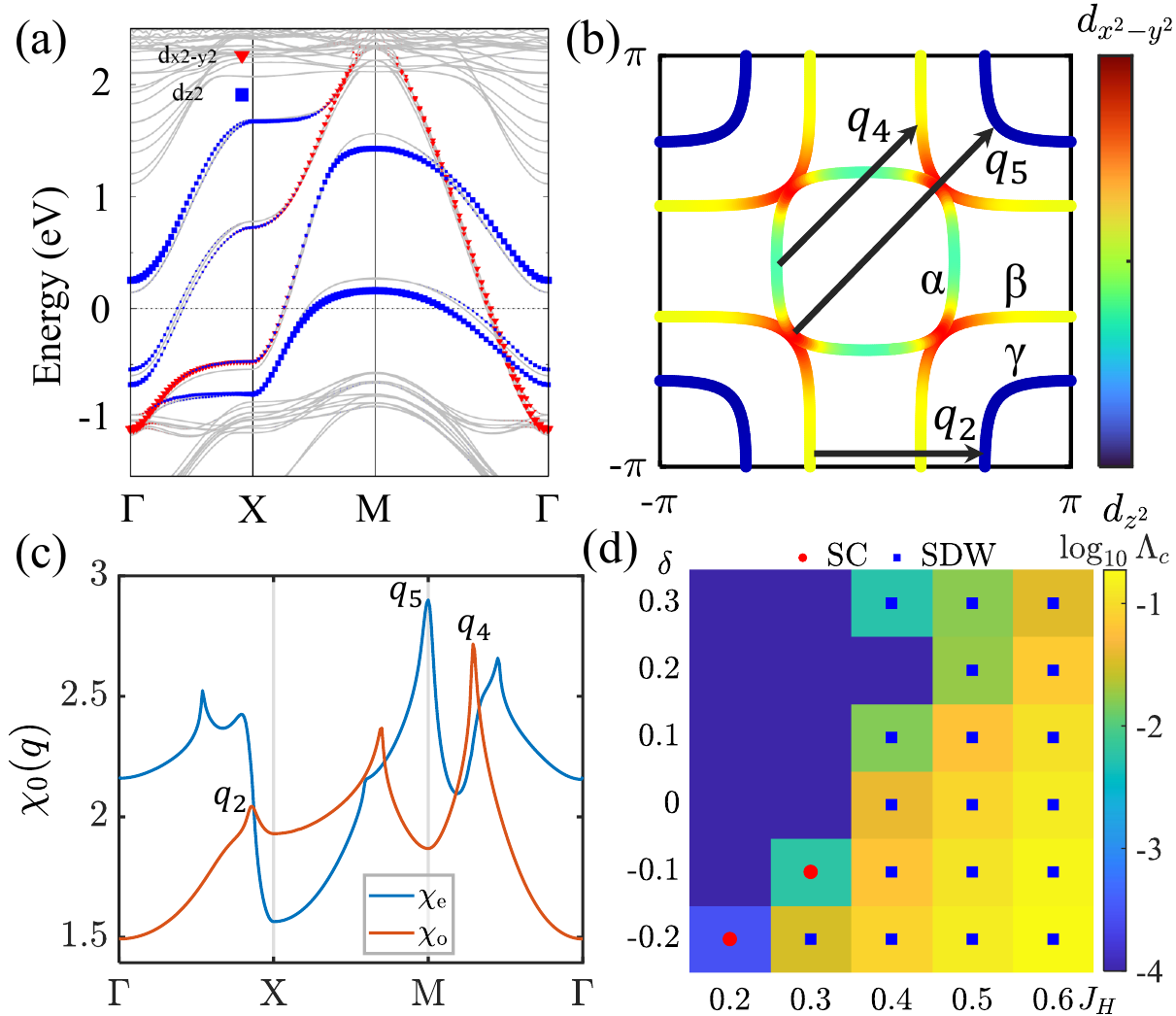}}
\caption{
(a) {\it ab initio} band structure of stretched LNO thin films with an in-plane constant $a=3.90 \text{\r{A}}$ and corresponding Fermi surfaces (b) at pristine filling $n=3.0$.  (c) Bare spin susceptibility in even and odd channels. (d) Phase diagram of instabilities as a function of Hund's coupling $J_H$ and hole doping $\delta$ at fixed $U=3$ eV. Symbols and color scheme are consistent with Fig.\ref{fig3} and unmarked regions indicate no instability above the critical cutoff of $10^{-4}$.
\label{fig4}}
\end{figure}

{\it Enhanced SDW instability in stretched LNO thin films}.
We further investigate the instability of LNO thin films under tensile strain. Experimental results indicate that LNO films on SrTiO\(_3\) do not exhibit superconductivity but rather show insulating behavior~\cite{bhatt2025resolvingstructuraloriginssuperconductivity}, despite theoretical calculations suggesting that their Fermi surfaces are similar to those of the bulk under pressure~\cite{Geisler2024,2024arXiv241204391Z}. Under tensile strain, LNO exhibit orthorhombic distortion~\cite{bhatt2025resolvingstructuraloriginssuperconductivity}, but its electronic structure remains similar to the tetragonal phase (see SM). Here, we present the band structure~(Fig. \ref{fig4}(a)) for the tetragonal phase with using an in-plane lattice constant of \(a=3.90\) \AA~, consistent with previous results~\cite{Geisler2024,2024arXiv241204391Z}. The corresponding Fermi surfaces from an effective TB model (Fig. \ref{fig4}(b)) show all three pockets becoming more rectangular. The $\alpha$ pocket gains increased $d_{z^2}$ character, and the \(d_{x^2-y^2}\) contribution to the \(\beta\) pocket  is also enhanced, leading to strong Fermi surface nesting at $\bm{q}_2$ and $\bm{q}_4$. Intriguingly, the rectangular \(\alpha\) pocket aligns well with the \(\gamma\) pocket when shifted by \(\bm{q}_5=(\pi,\pi)\), indicating strong Fermi surface nesting between them.  
From the susceptibility at \(\delta = 0\) shown in Fig.~\ref{fig4}(c), we observe a dominant sharp nesting peak around \(\bm{q}_5\) in the even channel, with weaker peaks at \(\bm{q}_4\) and \(\bm{q}_2\) in the odd channel. Electron doping can suppress the nesting at $\bm{q}_{5}$.
Due to multiple competing spin fluctuations, the undoped system does not develop instability for a low \(J_H/U < 0.1\) above the cutoff of \(10^{-4}\), as shown in Fig.~\ref{fig4}(d). However, when Hund's coupling is increased, the system exhibits a robust SDW instability with a wavevector of \(\bm{q}_4\), regardless of hole or electron doping.
The nesting at $\bm{q}_5$ involves two bonding pockets and induces interlayer ferromagnetic coupling, which suppresses interlayer pairing. This, together with  $\beta-\beta$ and $\gamma-\gamma$ nesting, renders superconductivity mostly subdominant. The absence of superconductivity and robust SDW seem to be consistent with the experimental observation of insulating behavior in the resistivity of stretched LNO thin films~\cite{bhatt2025resolvingstructuraloriginssuperconductivity}.

{\it Discussion and conclusion}. Our results reveal a unique superconducting mechanism in bilayer nickelates, governed by scatterings between Fermi surfaces with opposite mirror parity. Unlike iron based superconductors, where different SDW fluctuations generate competing pairings~\cite{Hirschfeld_2011}, in LNO two SDW fluctuations cooperatively enhance interlayer $s_{\pm}$-wave pairing through shared interlayer AFM coupling. In contrast, SDW fluctuations from the same-parity nesting are pairing-breaking. When pressure or charge doping causes the \(d_{z^2}\) bonding band to become incipient or cross \(E_\text{F}\), competing magnetic fluctuations destabilize SDW instability and superconductivity emerges. This mechanism offers a unified scenario for superconductivity in both bulk and thin-film LNO.
Once the bonding \(\gamma\) pocket fully appears, it introduces both pair-forming and pair-breaking effects. Our calculations show that this suppression dominates beyond the Lifshitz transition, producing a \(T_c\) maximum near the transition (Fig.~\ref{fig0}), which is consistent with the superconducting dome in Sr-doped LNO thin films~\cite{hao2025superconductivity}.

The maximum \(T_c\) reported in strained thin films is about 50 K~\cite{liu2025superconductivitynormalstatetransportcompressively, zhou2024ambientpressuresuperconductivityonset40}, well below the bulk under pressure. While improved sample quality may raise this value, our results suggest \(T_c\) in strained films will remain lower because compressive strain weakens the interlayer AFM coupling and thus interlayer pairing. External pressure along c axis should counteract this effect, strengthening interlayer coupling and increasing \(T_c\), analogous to infinite-layer nickelates~\cite{wang2022}. Under tensile strain, strong pair-breaking nesting suppresses superconductivity; however, charge doping can weaken this effect and may enable superconductivity once the SDW is suppressed.

In summary, we establish a direct link between ordering tendencies and Fermiology in bilayer nickelates: scattering between Fermi surfaces of different mirror parities governs SDW and superconductivity. Robust \(s_{\pm}\)-wave superconductivity can emerge when the \(d_{z^2}\) bonding band becomes incipient or crosses \(E_\text{F}\). These results provide a unified framework for superconductivity in both pressurized bulk and strained thin films.

{\it Acknowledgments}. We acknowledge the supports by National Natural Science Foundation of China (No. 12494594, No.11920101005, No. 11888101, No. 12047503, No. 12322405, No. 12104450), the Ministry of Science and Technology (Grant No. 2022YFA1403901), and  the New Cornerstone Investigator Program. X.W. is supported by the National Key R\&D Program of China (Grant No. 2023YFA1407300) and the National Natural Science Foundation of China (Grants No. No. 12574151, 12447103 and 12447101). C.C.L. is supported by the RIKEN TRIP initiative (RIKEN Quantum).

\bibliography{references_new250805}
\bibliographystyle{apsrev4-1}

\clearpage

\appendix
\begin{figure}[b]
\centerline{\includegraphics[width=0.5\textwidth]{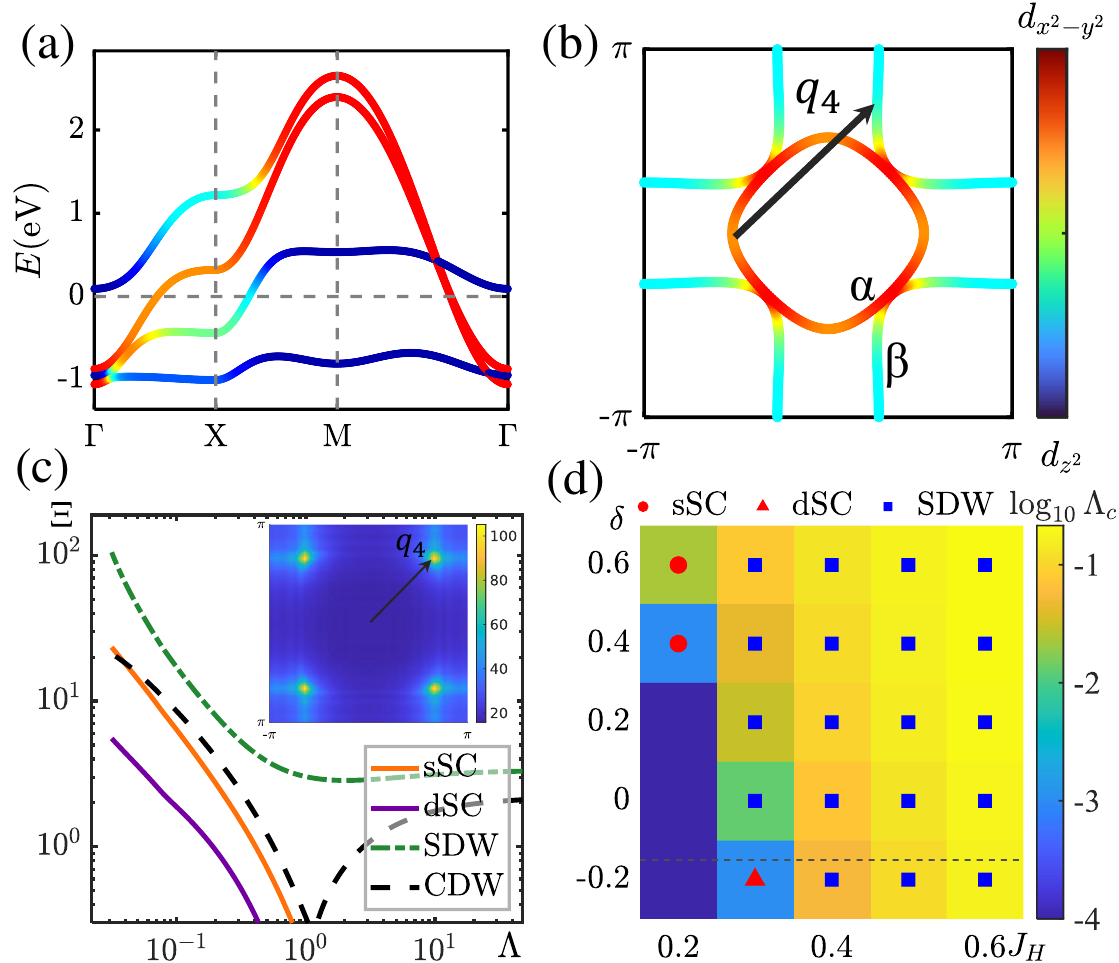}}
\caption{(a)
Tight-binding model band structure of interfacial NiO$_2$ bilayer and (b) corresponding Fermi surface at filling $n=3.34$.  (c) Typical FRG flows of leading eigenvalues in SC, SDW and CDW channels with $U=3$ eV, $J=0.1U$ and hole doping $\delta=0.2$. The inset figure shows the SDW leading eigenvalue distributions in BZ with the wave vectors of SDW given by the four symmetry related peaks along $\Gamma-M$ directions. (d) Phase diagram of instability as a function of Hund's coupling strength $J_H$ and hole doping level $\delta$ for a fixed $U=3$ eV. Symbols and color scheme are consistent with Fig.\ref{fig3} and unmarked regions indicate no instability above the critical cutoff of $10^{-4}$. The black dashed line denotes the Lifshitz transition point $\delta_0=0.15$ with the appearance of electron pocket around the $\Gamma$ point for $\delta>\delta_0$. \label{fig2}}
\end{figure}

{\it Instability of interfacial NiO$_2$ bilayer}. We examine the electronic instability of the NiO\(_2\) bilayer at the interface as a function of doping. Fig. \ref{fig2}(a) presents the orbital-resolved band structure from the  effective TB model for this bilayer, and Fig. \ref{fig2}(b) illustrates the corresponding Fermi surfaces. The onsite energy difference for the \(d_{z^2}\) orbital between the top and bottom layers is approximately 0.23 eV according to Wannier fitting. Compared to the pressurized bulk case, there is an increase in the weight of both the \(d_{x^2-y^2}\) bonding orbital on the \(\alpha\) pocket and the \(d_{z^2}\) antibonding orbital on the \(\beta\) pocket. This occurs as the \(d_{z^2}\) antibonding orbital moves away from the Fermi level, while the antibonding orbital  shifts closer to it.
In Fig.~\ref{fig2}(c), we depict the representative FRG flow for a slightly hole-doped case with a reasonable interaction setting of \(U=3\) eV and \(J/U=0.1\). The  SDW state is dominant from the outset and diverges at a relatively high cutoff, indicating a strong SDW instability, while superconductivity instability is weak. The inset shows the leading eigenvalues in the 2D Brillouin zone, with divergence occurring at $\bm{q}_4\approx(3\pi/4,3\pi/4)$, corresponding to Fermi surface nesting between the \(\alpha\) and \(\beta\) pockets.
To simulate charge doping in realistic systems~\cite{zhou2024ambientpressuresuperconductivityonset40}, we performed systematic calculations for both hole-doped ($\delta>0$) and electron-doped ($\delta<0$)  cases for various Hund's coupling. The resulting phase diagram, shown in Fig.~\ref{fig2}(d), demonstrates a clear dominance of the SDW state with a wavevector on the $\Gamma$-M line across a wide doping regime. In contrast, superconductivity only appears within a narrow range of weak Hund's coupling with small critical cutoffs. Notably, the leading pairing is \(d\)-wave for the electron-doped case after the Lifshitz transition (denoted by the black dashed line), where an electron pocket around the \(\Gamma\) point arises from the \(d_{z^2}\) antibonding orbital.
Based on these findings, we deduce that the observed superconductivity is unlikely to originate from the interfacial NiO$_2$ bilayer.

\begin{figure}[t]
\centerline{\includegraphics[width=0.5\textwidth]{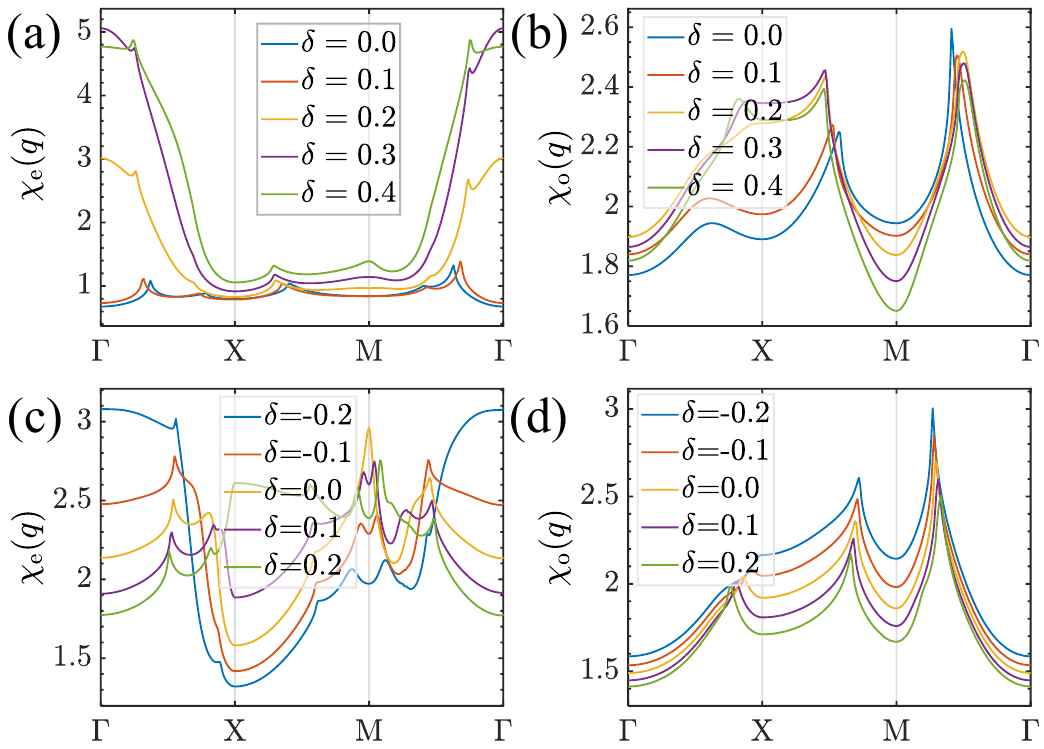}}
\caption{Bare spin susceptibilities along the high-symmetry momentum path \(\Gamma\)-\(X\)-\(M\) in the even (\(\chi_{\mathrm{e}}\)) and odd (\(\chi_{\mathrm{o}}\)) channels for the third NiO\(_2\) bilayer under compressive strain (a,b) and tensile strain (c,d) at various hole doping levels \(\delta\). \label{fig333}}
\end{figure}

{\it Doping dependent susceptibility of compressed and stretched LNO thin films}. To compare the spin fluctuations arising from different electronic structures in LNO thin films under compressive and tensile strains, Fig.~\ref{fig333} shows the bare spin susceptibilities in the even and odd channels for the third NiO\(_2\) bilayer under compressive strain and for a stretched LNO thin film. For the third NiO\(_2\) bilayer under compressive strain, the even-channel spin fluctuations are relatively weak, whereas the odd-channel fluctuations are stronger due to the good nesting between the $\alpha$ and $\beta$ pockets in the undoped and lightly hole-doped cases. The strong odd-channel fluctuations, characterized by interlayer antiferromagnetic correlations, can drive an SDW instability at $\mathbf{q}_4$ once intermediate interactions are included. With hole doping levels approaching the Lifshitz transition, the $\mathbf{q}_4$ nesting is weakened, while additional odd-channel fluctuations near the $X$ point emerge due to nesting between the bonding $\beta$ and antibonding $\gamma$ bands. These two types of strong odd-channel fluctuations with interlayer antiferromagnetic character can cooperate to enhance interlayer spin-singlet pairing, corresponding to the bonding–antibonding $s_{\pm}$ state in band space. With further hole doping beyond the Lifshitz transition near $\delta_c \approx 0.22$, the even-channel fluctuations grow due to the flatness of the $d_{z^2}$ bonding band near the M point. In this regime, the odd-channel fluctuations associated with interlayer ferromagnetic coupling act as pair-breaking effects for interlayer spin-singlet pairing, thereby explaining the decline of $T_c$ for $s_{\pm}$ pairing after the Lifshitz transition, as odd-channel pair-breaking fluctuations are enhanced.  
For the stretched LNO thin film with $a=3.90$\AA~, the even-channel fluctuations remain consistently stronger than those in the odd channel for both hole ($\delta>0$) and weak electron ($\delta<0$) dopings. In particular, the strong nesting between the bonding $\alpha$ and $\gamma$ pockets strongly enhances the even-channel fluctuations near the M point. The strong odd-channel pair-breaking fluctuations for interlayer pairing account for the absence of superconductivity in the tensile-strained thin film. However, at larger electron dopings where the $\gamma$ pocket becomes smaller, the even-channel fluctuations can be suppressed, which may be beneficial to emergence of superconductivity.

\end{document}